\documentclass[12pt]{article}

\usepackage{arxiv}
\usepackage{graphicx}
\usepackage{multirow}
\usepackage{titlesec}
\usepackage{balance} 
\usepackage{stfloats}
\usepackage{amssymb,amsmath,amsfonts,eurosym,ulem,graphicx, caption, color,setspace, comment, footmisc,caption, pdflscape, subfigure, array, hyperref, algpseudocode}
\usepackage{xcolor}
\usepackage{amsmath}
\usepackage{caption}
\usepackage{amsthm}
\usepackage{lipsum}
\usepackage{etoolbox}
\usepackage{enumitem}
\usepackage{float}
\usepackage{physics}
\allowdisplaybreaks[3]
\usepackage{booktabs}
\usepackage{threeparttable}
\usepackage{soul}
\usepackage{amsmath}
\usepackage{graphicx}
\usepackage{multirow}
\usepackage{tabularx}
\usepackage{array}
\usepackage[most]{tcolorbox}
\usepackage{parskip}
\usepackage[ruled]{algorithm2e}
\newtcbtheorem{Summary}{\bfseries Research Question}{enhanced,drop shadow={black!50!white},
  coltitle=black,
  top=0.2in,
  attach boxed title to top left=
  {xshift=1.5em,yshift=-\tcboxedtitleheight/2},
  boxed title style={size=small, colback = pink}
}{summary}

\def\BibTeX{{\rm B\kern-.05em{\sc i\kern-.025em b}\kern-.08em
    T\kern-.1667em\lower.7ex\hbox{E}\kern-.125emX}}
    
\usepackage{amsmath}
\usepackage{caption}
\usepackage{amsthm}
\usepackage{lipsum}
\usepackage{etoolbox}
\usepackage{enumitem}
\usepackage{float}
\usepackage{physics}
\allowdisplaybreaks[3]
\usepackage{geometry}
\usepackage{booktabs}
\usepackage{threeparttable}
\usepackage{soul}
\usepackage{amsmath}
\usepackage{graphicx}
\usepackage{multirow}
\usepackage{tabularx}
\usepackage{array}

\renewcommand{\shorttitle}{A Quantitative Analysis of OSS Quality: Insights from Metric Distributions}

\title{\textbf{\Large Software Code Quality Measurement: Implications from Metric Distributions \\}}

\author{Siyuan Jin$^{1,2}$, Mianmian Zhang$^{1}$, Yekai Guo$^{3}$, Yuejiang He$^{1}$, \\
\textbf{Ziyuan Li$^{4,1,*}$, Bichao Chen$^{1,*}$, Bing Zhu$^{1,*}$, and Yong Xia$^{1,*}$}\\
	\normalsize $^{1}$ HSBC Laboratory, Guangzhou, China\\
	\normalsize $^{2}$ School of Business and Management, Hong Kong University of Science and Technology, Hong Kong, China\\
	\normalsize $^{3}$ School of Data Science, Fudan University, Shanghai, China\\
 	\normalsize $^{4}$ School of Physics, Sun Yat-sen University, Guangzhou, China\\
	\normalsize liziyuan3@mail.sysu.edu.cn, bichao.chen@hsbc.com, bing1.zhu@hsbc.com, yong.xia@hsbc.com\\
	\normalsize *corresponding author
}

\newcolumntype{C}{>{\centering\arraybackslash}X}
\begin{document}
\maketitle

\begin{abstract}
Software code quality is a construct with three dimensions: maintainability, reliability, and functionality. Although many firms have incorporated code quality metrics in their operations, evaluating these metrics still lacks consistent standards. We categorized distinct metrics into two types: 1) monotonic metrics that consistently influence code quality; and 2) non-monotonic metrics that lack a consistent relationship with code quality. To consistently evaluate them, we proposed a distribution-based method to get metric scores. Our empirical analysis includes 36,460 high-quality open-source software (OSS) repositories and their raw metrics from SonarQube\footnote{https://www.sonarsource.com} and CK\footnote{https://github.com/mauricioaniche/ck}. The evaluated scores demonstrate great explainability on software adoption. Our work contributes to the multi-dimensional construct of code quality and its metric measurements, which provides practical implications for consistent measurements on both monotonic and non-monotonic metrics. 
\end{abstract}
\vspace{1.5ex}

\onehalfspacing
\section{Introduction}
Code quality refers to the extent to which code is well-written and meets given needs \cite{lee2009measuring}. Precise code quality measurement can improve software products, increase user satisfaction, and save costs of IT systems \cite{kekre1995drivers}, which influences the software adoption \cite{crowston2003defining, levine2010quality}. 
Therefore, numerous firms have incorporated measurements to evaluate code quality. However, these methods display a wide range of diversity and lack consistent standards. 

\begin{figure}
    \centering
    \includegraphics[width= 0.5 \textwidth]{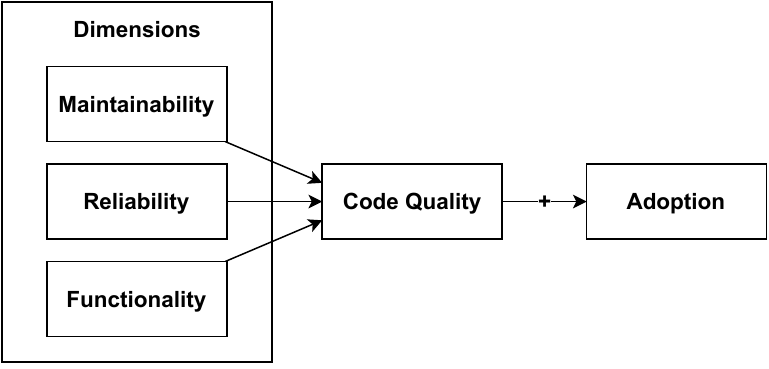}
    \caption{Multi-Dimensional Construct}
    \label{fig: Construct}
\end{figure}

Figure \ref{fig: Construct} shows that code quality is a multi-dimensional construct that includes dimensions: maintainability, reliability, and functionality \cite{lee2009measuring}. Based on the literature on code quality dimension measurements, we identified 20 distinct metrics and divided them into monotonic and non-monotonic metrics. Monotonic metrics consistently impact code quality, while non-monotonic metrics lack a consistent relationship with code quality (Figure \ref{fig:metric_dist}). Most monotonic metrics exhibit a monotonically decreasing relationship with code quality. A case in point is the number of code smells, which, when it rises, usually denotes a corresponding decline in the overall code quality.

The literature remains a gap in the methodologies for consistently evaluating both types of code quality metrics, especially non-monotonic metrics. Therefore, the most prevalent method for assessing code quality within firms continues to be peer code review \cite{sadowski_modern_2018}. To consistently evaluate both types of code quality metrics, we propose a distribution-based method, which shows great explainability on software adoption.

\begin{Summary}{}{firstsummary}
How to consistently evaluate both monotonic and non-monotonic metrics for software code quality?
\end{Summary}
We evaluated metric scores by analyzing their probability distributions among high-star OSS. For monotonic metrics, we fit an exponential distribution and use the weighted distance from threshold parameters in their cumulative distribution functions (CDFs) as their scores. For non-monotonic metrics, we fit an asymmetric Gaussian distribution and use the weighted distance away from the central point in their CDFs as their scores. The evaluated scores range from $0\sim100$ for each metric.

We conducted our empirical analysis on 36,460 GitHub OSS repositories. The selection of repositories with a high number of stars results in a more rigorous evaluation as those higher-quality repositories are used as reference points. The repositories that are slightly worse than our selected ones typically receive extremely low scores due to our sharper distributions from high-quality repositories.

\begin{Summary}{}{secondsummary}
What are the implications of the evaluated scores on software adoption?
\end{Summary}


We investigated the explainability of our code quality metric scores on OSS stars. The number of stars reflects OSS quality and adoption \cite{medappa2019does}. With standard machine learning approaches, we use R-squared (R2) and accuracy as measures to assess their explanatory power. The results show our code quality scores can explain the number of OSS stars well. Our methodology can be applied to different target variables, providing a flexible strategy in various contexts.

This work has the following contributions. Prior literature has discussed diverse code quality metrics \cite{mccabe1976complexity, stamelos_code_2002, shin2010evaluating} without consistent metric evaluations. We extended them by dividing code quality metrics into two types and evaluated them with a novel distribution-based method. We conducted our empirical analysis on 36,460 GitHub OSS repositories. We used our evaluated scores to explain the OSS adoption \cite{lee2009measuring}, generating implications into how code quality may influence the OSS adoption. Our study advances the understanding of code quality with two different types of code quality metrics and contributes to better quality control standards and practices.

\section{Literature Review} \label{sec: lr} 
Open Source Software (OSS) uses a communal approach to software development, which significantly increases their code quality \cite{ljungberg2000open, von2006promise} and fuels innovation \cite{levine2014open, vitharana2010impact, kumar2011competitive}. Many firms and developers actively contribute to and utilize OSS \cite{mehra2011firms}. These contributions serve to incrementally enhance the overall OSS code quality. The reuse of OSS has been widely adopted because many OSS have high code quality \cite{harter2000effects} which can effectively reduce the search costs for developers \cite{haefliger2008code, sojer2010code}. Therefore, we use OSS as the benchmark to evaluate software code quality.

Many research studied performance evaluations for software \cite{al2017empirical}. Inappropriate performance measurements have been identified as a major cause of IT systems failing \cite{kekre1995drivers, fitoussi2012outsourcing}. As new technologies and techniques emerge, such as blockchain systems \cite{jin2022cev}, and AI agent systems \cite{diederich2022design}, more precise measurements of software code quality are needed \cite{anderson2019evidence}. Our approach sets itself apart from past studies by considering the distribution of high-quality software and delivering accurate scores for each software.

Code quality encompasses various dimensions \cite{polites2012conceptualizing}. The IEEE standard defines code quality as the collective features and characteristics of software that meet given needs \cite{fitzpatrick1996software}. Later on, user-friendliness and useful functionalities are included in the definition of code quality \cite{lee2009measuring}, echoing the three dimensions in the ISO/IEC 25010 standard \cite{klima2022selected}: maintainability, reliability, and functionality \cite{athanasiou2014test}. Similarly, other studies have similar dimensions: maintainability \cite{motogna2023empirical}, readability \cite{gonzalez2023reliability}, and functionality \cite{shen2020api}. We base on the literature to define the construct and dimensions in Table \ref{tb: construct}.

\begin{table*}[]
\small
\renewcommand{\arraystretch}{1.2}
\centering
\caption{Construct Definition \label{tb: construct}}
\begin{tabular}{@{}p{1.5cm}p{4cm}p{2.2cm}p{8cm}@{}}
\toprule
\textbf{Construct} & \textbf{Definition} & \textbf{Dimensions} & \textbf{Definition} \\ 
\midrule
\multirow{3}{5cm}{Code Quality} & \multirow{3}{4.5cm}{The extent to which code is well-written and meets given needs. \cite{lee2009measuring}} & Maintainability & The code is easy to understand, enhance, or correct. \cite{deligiannis2003empirical} \\
& & Reliability & The code is user-friendly and stable. \cite{lee2009measuring} \\
& & Functionality & The code has useful functions. \cite{lee2009measuring} \\ 
\bottomrule
\end{tabular}
\end{table*}

Code quality has metrics, including the size of components \cite{stamelos_code_2002}, code complexity \cite{mccabe1976complexity,shin2010evaluating}, and so on. However, most existing metric identifications have focused on monotonic metrics rather than non-monotonic metrics, because monotonic metrics have a consistent relationship with software code quality. Our paper considers both types and proposes a uniform solution for evaluating them.

Reflective measurements, such as the number of stars \cite{medappa2019does}, can indicate the overall level of software code quality. Although OSS adoption activities are determined by many factors, such as commitment \cite{maruping2019developer}, transparency \cite{shaikh2016folding}, and leader resources \cite{dong2021project}, OSS adoption decision can reflect good OSS code quality. Lee et al. \cite{lee2009measuring} highlight the impact of code quality on user satisfaction and adoption. The OSS repositories that see the highest adoption rate are often those that maintain exceptional code quality. Therefore, we use GitHub stars as a reflective measure of code quality.

\section{Methodologies} \label{sec: method}
Our study employs the number of stars as a reflective measurement for identifying good-quality repositories. We divide code quality metrics into two distinct groups. We then analyze various code quality metric distributions and introduce a consistent distribution-based approach to evaluate all metrics within these categories.

We first map out the distribution of each metric in high-star OSS repositories and then score them according to their corresponding metric CDFs.

\begin{figure*}
\centering
\includegraphics[width=0.8\textwidth]{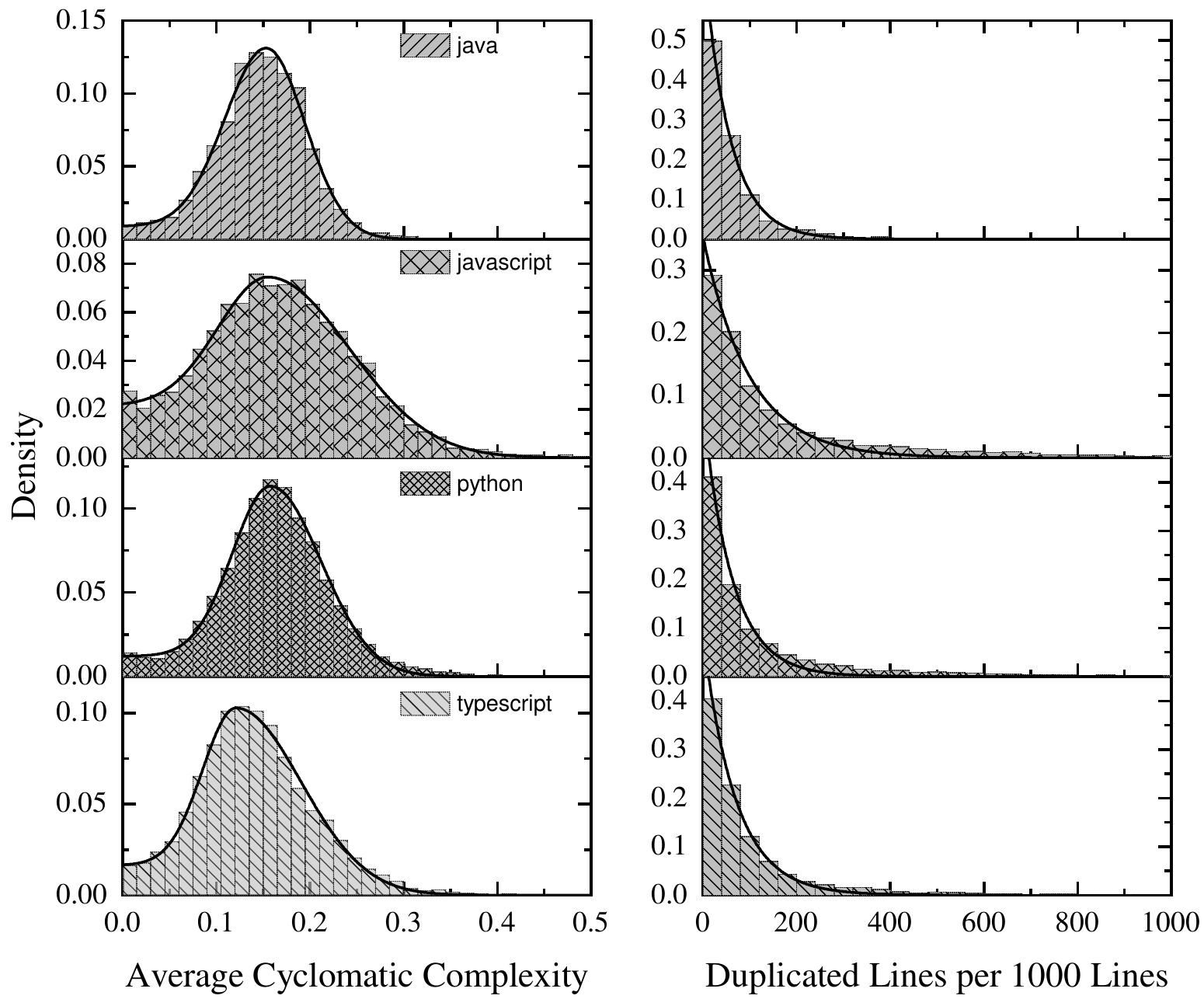}
\caption{Examples of Non-Monotonic Metric Distribution and Monotonic Metric Distribution}
\label{fig:metric_dist}
\end{figure*}

Table \ref{tab:metric_definition} presents two different types of metric distributions: monotonic and non-monotonic metrics. We fit exponential distributions to monotonic metrics, the probability distribution function (PDF) of which reads as:
\begin{equation}\label{eq:exp_pdf}
f_1(x;c,\lambda) = \begin{cases}
0 & \text{if } x \leq c  \\
\lambda \exp\left[-\lambda {(x-c)}\right] & \text{if } x > c 
\end{cases} \
\end{equation}
where $\lambda$ and $c$ are the fitting parameters. The corresponding score function based on the CDF of Eq. \eqref{eq:exp_pdf} reads as
\begin{equation}\label{eq:exp_score}
\small
\begin{split}
& M_1(x; c, \lambda) =  100 \times \begin{cases}
1 & \text{if } x \leq c \\
\exp\left[-\lambda {(x-c)}\right]  & \text{if } x > c
\end{cases} \
\end{split}
\end{equation}
The score falls into the range of $0\sim100$ and it peaks at $c$ and decays exponentially for $x>c$.

The non-monotonic metrics follow an asymmetric Gaussian distribution (see the left of Fig. \ref{fig:metric_dist}), the PDF of which reads as
\begin{equation}\label{eq:agass_pdf}
\begin{split}
& f_2(x;\mu, \sigma_1, \sigma_2) = \\
&
\begin{cases}
\frac{1}{\sqrt{2\pi}}\frac{2}{{\sigma_1} + {\sigma_2}} \exp\left(-\frac{(x-\mu)^2}{2\sigma_1^2}\right) & \text{if } 0\leq x < \mu \\
\frac{1}{\sqrt{2\pi}}\frac{2}{{\sigma_1} + {\sigma_2}} \exp\left(-\frac{(x-\mu)^2}{2\sigma_2^2}\right) & \text{if }  x \geq \mu 
\end{cases} \
\end{split} 
\end{equation}
where $\mu, c, \sigma_1,\sigma_2$ are fitting parameters representing the peak position, peak height on the right, and peak widths on each side, respectively. The corresponding score function is
\begin{equation}\label{eq:agass_score}
\begin{split}
M_2(x, \mu, \sigma_1, \sigma_2) =  100 \times
\begin{cases}
1 - \operatorname{erf}\left(\frac{x-\mu}{\sigma_1 \sqrt{2}}\right) & \text{if }  0\leq x < \mu\\
1 - \operatorname{erf}\left(\frac{x-\mu}{\sigma_2 \sqrt{2}}\right) & \text{if } x \geq \mu
\end{cases}\
\end{split} 
\end{equation}
where the score falls into the range of $0\sim100$, peaks at $\mu$, and decays according to the Z-score of the Gaussian function on each side.

To obtain an overall score, we assign weights to individual scores. The overall score for a given repository, denoted by $k$, can be computed as follows:
\begin{equation} \label{eq: overall_score}
Q^{overall}_{k} = \sum_{i}{{\omega_{i}} \cdot Q^{metric}_{i,k}} \, ,
\text{subject to:} \sum_i{\omega_i} = 1 .
\end{equation}
The weights $\omega_i$ are derived from the importance values from supervised learning models for metric scores to a target variable such as repository stars.

\begin{figure*}
\centering
\includegraphics[width = 0.95\textwidth]{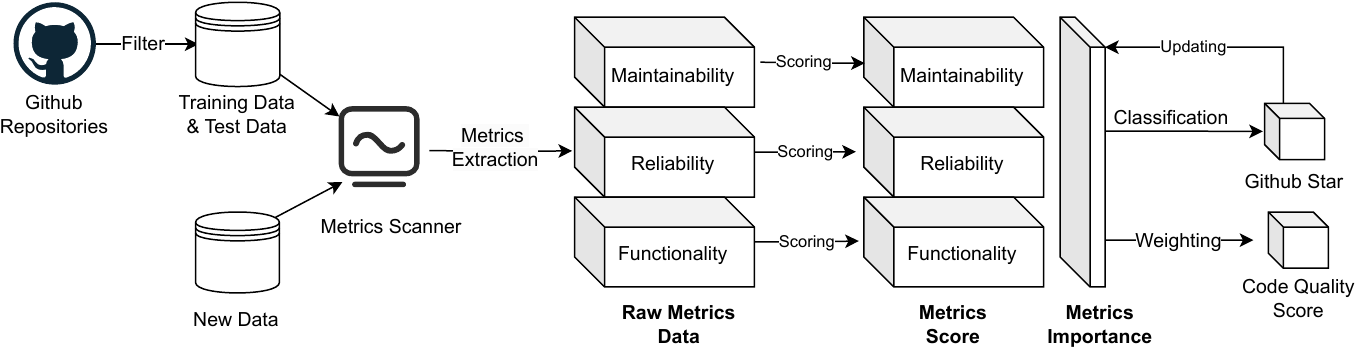}
\caption{Workflow for Code Quality Scoring with GitHub Stars as the Target Variable}
\label{fig: Workflow}
\end{figure*}

\section{Empirical Analysis} \label{sec: application}

\subsection{Data Sources}
\begin{table*}
\renewcommand{\arraystretch}{1.2}
\small
\centering
\caption{Statistical Summary}
\begin{tabular}{@{}ccccc@{}}
\toprule
\textbf{Programming Language} &  \textbf{Max Number of Stars} & \textbf{Min Number of Stars} & \textbf{Number of Filtered Repositories} \\ \midrule
Java                                        & 50k                     & 100                     & 1,645                           \\
Python                                      & 228k                    & 260                     & 16,096                          \\
JavaScript                                & 107k                    & 270                     & 7,722                           \\
TypeScript                                  & 202k                    & 60                      & 10,997                          \\ \bottomrule
\end{tabular}
\label{tab:repo-statistics}
\end{table*}

GitHub is the largest OSS management platform that has more than 39 million public repositories (As of June 2023). We selected a subset of repositories with Java, Python, JavaScript, and TypeScript as the main programming languages and sorted them by the number of stars. We collected code from the top $\sim20,000$ repositories for each programming language. The number of GitHub stars is a measure of OSS adoption \cite{medappa2019does}. We removed non-engineering repositories by pattern matching, such as a guide for Java interviews in \href{https://github.com/Snailclimb/JavaGuide}{JavaGuide}. 

We used code scanners to obtain metrics. Scripting language repositories (Python, Javascript, TypeScript) can be directly imported, while non-scripting Java repositories need to be compiled first. Compiling Java repositories is challenging due to their different JDK, maven, or Gradle versions. Therefore, we only chose repositories with GitHub releases for compilation, which led to 36,460 repositories and over 600 million lines of code. Table ~\ref{tab:repo-statistics} reports the statistics of the cloned repositories. The minimum number of repository stars is above 50, which demonstrates good code quality compared to overall OSS repositories.

\subsection{Metrics Overview}
We used SonarQube and CK to extract metrics from OSS repositories. For Java repositories, we generated over 100 metrics and selected 20 based on the ISO/IEC 25010 international standard \cite{klima2022selected}. For scripting language repositories, we only extracted 12 metrics. Table~\ref{tab:metric_definition} shows the 20 metrics with their corresponding ISO/IEC 25010 characteristics. 

\begin{table*}
\centering
\small
\renewcommand{\arraystretch}{1.2}
\begin{threeparttable}
\caption{Definition of 20 Code Quality Metrics}
\begin{tabular}{@{}p{2cm}p{4cm}p{10cm}@{}}
\toprule
\textbf{Dimension}                         & \textbf{Metric}                                                             & \textbf{Definition}                                                                                                                    \\ \midrule
\multirow{12}{*}{Maintainability} & Cyclomatic Complexity\tnote{a}                    & Number of independent paths through code.                                                                                     \\
                                  & File Complexity\tnote{b} & Cyclomatic complexity averaged by files.                                                                                      \\
                                  & Cognitive Complexity\tnote{a}                     & Combination of cyclomatic complexity and human assessment.                                                                    \\
                                  & Code Smells\tnote{a}                              & Number of code smell issues.                                                                                                  \\
                                  & Coupling Between Objects                                           & Number of classes coupled to a particular class.                                                                     \\
                                  & Fan-in                                                             & Number of input dependencies a class has.                                                                                     \\
                                  & Fan-out                                                            & Number of output dependencies a class has.                                                                                    \\
                                  & Depth Inheritance Tree                                             & Number of "fathers" a class has.                                                                                    \\
                                  & Number of Children                                                 & Number of immediate subclasses that a particular class has.                                                                   \\
                                  & Lack of Cohesion of Methods                                        & Degree to which class methods are coupled.                                                                                    \\
                                  & Tight Class Cohesion                                               & Ratio of the number of pairs of directly related methods in a class to the maximum number of possible methods in the class.   \\
                                  & Loose Class Cohesion                                               & Ratio of the number of directly or indirectly related method pairs in a class to the maximum number of possible method pairs. \\ \midrule
\multirow{3}{*}{Reliability}      & Total Violations\tnote{a}                         & Number of issues including all severity levels.                                      \\
                                  & Critical Violations\tnote{a}                      & Number of issues of the critical severity.                                                                                    \\
                                  & Info Violations\tnote{a}                          & Number of issues of the info severity.                                                                                        \\ \midrule
\multirow{5}{*}{Functionality}    & Line to Cover\tnote{a}                            & Lines to be covered by unit tests.                                                                                                  \\
                                  & Comment Lines\tnote{b}                            & Number of comment lines.                                                                                                      \\
                                  & Duplicated Blocks\tnote{e}                        & Number of duplicated blocks of line.                                                                                          \\
                                  & Duplicated Files\tnote{d}                         & Number of files involved in duplicated blocks.                                                                                \\
                                  & Duplicated Lines\tnote{c}                         & Number of lines involved in duplicated blocks.                                                                                \\ \bottomrule
\end{tabular}
\begin{tablenotes}
\small
\item[a] Normalized by Non-comment Line of Codes.
\item[b] Normalized by Sum of Non-comment Line of Codes and Comment Lines.
\item[c] Normalized by Line of Codes.
\item[d] Normalized by Number of Files.
\item[e] Normalized by Number of Statements.
\end{tablenotes}
\label{tab:metric_definition}
\end{threeparttable}
\end{table*}

We normalized metrics to ensure score fairness. Cyclomatic complexity, cognitive complexity, code smells, line to cover, and violations-related metrics are normalized by non-comment lines of code, duplicated lines are normalized by lines of code, and comment lines are normalized by the sum of non-comment lines and comment lines, to account for repository size. File complexity and duplicated files are normalized by the number of files, and duplicated blocks are normalized by the number of statements to adjust for differences across repositories. This normalization process results in a more unbiased score for the metrics across different OSS.

\subsection{Importance Weights}
We use standard machine-learning approaches to derive weights for different metric scores and calculate a repository's overall code quality score. Our model can explain OSS adoption (Github stars) using evaluated scores.

Figure~\ref{fig: Workflow} illustrates the entire process from data collection to final scores. We use custom data filters to ensure genuine engineering repositories are retained. We extract code quality metrics using a metric scanner and generate metric scores using the distribution-based method in Section~\ref{sec: method}, with each programming language having its distribution for each metric. We implement a Gradient Boosting Classifier (GBC) model with 0-1 labels as dependent variables based on the number of GitHub stars. We label the top and bottom quintiles (20\%) of the OSS repository stars as 1 and 0, respectively. The model generates importance values as weights for each metric. Finally, we obtain a weighted average code quality score according to Eq. \eqref{eq: overall_score}.

\begin{algorithm}
\small
  \SetAlgoLined
  \KwIn{Training dataset $\mathcal{D} = \{(\mathbf{m}_i, c_i)\}_{i=1}^N$, number of iterations $T$}
  \KwOut{Ensemble model $F(\mathbf{m})$}
  
  Initialize model $F_0(\mathbf{m}) = 0$\;
  
  \For{$t=1$ to $T$}{
    Compute the negative gradient: $r_{it} = -\frac{\partial L(c_i, F(\mathbf{m}_i))}{\partial F(\mathbf{m}_i)} \bigg|_{F(\mathbf{m}) = F_{t-1}(\mathbf{m})}$\;
    
    Fit a base learner $h_t(\mathbf{m})$ to the negative gradient: $h_t(\mathbf{m}) = \arg\min_h \sum_{i=1}^N L(c_i, F_{t-1}(\mathbf{m}_i) + h(\mathbf{m}_i))$\;
    
    Update the ensemble model: $F_t(\mathbf{m}) = F_{t-1}(\mathbf{m}) + \eta h_t(\mathbf{m})$, where $\eta$ is the learning rate\;
  }
  \caption{GBC in Our Context}
  \label{alg:gbc}
\end{algorithm}

The GBC algorithm is presented in Algorithm \ref{alg:gbc}, where each data point contains a metric score $\mathbf{m}_i$ and its corresponding classification $c_i$ according to its GitHub star. We divide the whole dataset into a training ($\mathcal{D}$) and a validation set by a ratio of 4:1. The GBC algorithm works with an ensemble model $F_0(\mathbf{m})$ and we fine-tune it by fitting base learners $h_t(\mathbf{m})$ to the loss function's negative gradient. The learning rate $\eta$ determines the base learners' contribution, resulting in the final ensemble model $F(\mathbf{m})$ providing the aggregate prediction.

\section{Results} \label{sec: result}

\subsection{Metric Distributions}

We conducted our empirical analysis on 36,460 GitHub OSS repositories. The selection of high-star repositories provides a more critical evaluation, because they generally have better performance, resulting in sharper distributions. 

Table~\ref{tab:metric-parameters1} and Table~\ref{tab:metric-parameters2} present the fitted parameters for the asymmetric Gaussian [Eq. \eqref{eq:agass_pdf}] and Exponential [Eq. \eqref{eq:exp_pdf}] distributions, respectively. Java repositories have 8 more maintainability metrics describing cohesion and coupling in the codes, which are absent for other programming languages due to a lack of proper metric scanners.  

\begin{table*}
\small
\centering
\renewcommand{\arraystretch}{1.2}
\caption{Parameters of the Fitted Asymmetric Gaussian Distributions ($\boldsymbol{\mu}$, $\boldsymbol{\sigma_1}$, $\boldsymbol{\sigma_2}$)}
\begin{tabular}{@{}ccccc@{}}
\toprule
{\textbf{Metric}} & Java$(\boldsymbol{\mu}$,$\boldsymbol{\sigma_1}$,$\boldsymbol{\sigma_2})$ & JavaScript$(\boldsymbol{\mu}$,$\boldsymbol{\sigma_1}$,$\boldsymbol{\sigma_2})$ & Python$(\boldsymbol{\mu}$,$\boldsymbol{\sigma_1}$,$\boldsymbol{\sigma_2})$ & TypeScript$(\boldsymbol{\mu}$,$\boldsymbol{\sigma_1}$,$\boldsymbol{\sigma_2})$ \\ \midrule
Cyclomatic Complexity            & (155.228,50.947,40.902)                                                      & (166.692,88.415,78.289)                                                       & (162.321,53.497,52.789)                                                    & (127.273,51.616,66.733)                                                        \\

Cognitive Complexity            &(50.870,40.120,75.664)                                                                         &(33.238,32.586,121.541)                                                    &(170.042,33.546,0.000)                                                    & (29.619,22.964,81.617)                                                      \\

Comment Lines            &(15.841,11.451,137.269)                                                                        &(0.007,6.575,96.312)                                                        &(91.730,64.805,148.192)                                                    & (0.002,9.300,72.443)                                                      \\
Fan-in                                   & (1.101,0.463,1.217)                                & /                                                & /                                            & /                                                \\
Fan-out                          & (5.181,2.043,4.639)                                                      & /                                                                              & /                                                                          & /                                                                              \\
Loose Class Cohesion             & (0.329,0.149,0.176)                                                      & /                                                                              & /                                                                          & /                                                                              \\
Tight Class Cohesion             & (0.228,0.100,0.128)                                                      & /                                                                              & /                                                                          & /                                                                              \\
Coupling Between Objects         & (7.055,2.580,5.086)                                                      & /                                                                              & /                                                                          & /                                                                              \\ \bottomrule
\end{tabular}
\label{tab:metric-parameters1}
\end{table*}

\begin{table*}
\centering
\small
\setlength{\tabcolsep}{10pt}
\renewcommand{\arraystretch}{1.2}
\caption{Parameters of the Fitted Exponential Distributions ($\boldsymbol{c}$, $\boldsymbol{\lambda}$)}
\begin{tabular}{@{}ccccc@{}}
\toprule
\textbf{Metric}                                   & Java$(\boldsymbol{c},\boldsymbol{\lambda})$ & JavaScript$(\boldsymbol{c},\boldsymbol{\lambda})$ & Python$(\boldsymbol{c},\boldsymbol{\lambda})$ & TypeScript$(\boldsymbol{c},\boldsymbol{\lambda})$ \\ \midrule
File Complexity & (0,0.485)                                  & (0,0.884)                                       & (0,0.917)                                   & (0,0.492)                                       \\
Code Smells                              & (1.123,50.731)                                 & (0.036,60.260)                                       & (0.004,37.177)                                   & (0.017,16.530)                                       \\

Depth Inheritance Tree                   & (1.003,0.502)                                  & /                                                & /                                            & /                                                \\
Number of Children                       & (0.002,0.137)                                  & /                                                & /                                            & /                                                \\
Lack of Cohesion of Methods              & (0.053,80.004)                                  & /                                                & /                                            & /                                                \\
Total Violations                         & (1.160,54.376)                                  & (0.054,63.313)                                       & (0.004,387.551177)                                   & (0.021,18.168)                                       \\
Critical Violations                      & (0.019,9.872)                                  & (0.020,48.811)                                       & (0.007,9.443)                                    & (0.005,5.497)                                        \\
Info Violations                          & (0.019,1.934)                                  & (0.001,1.436)                                        & (0.002,1.401)                                    & (0.003,1.535)                                        \\
Line to Cover                            & (0,0.000)                                  & (0,0.000)                                        & (0,0.000)                                    & (0,0.000)                                        \\
Duplicated Blocks                        & (0,0.015)                                  & (0.001,0.021)                                      & (0,0.010)                                    & (0,0.021)                                       \\
Duplicated Files                         & (0.003,0.135)                                  & (0.001,0.203)                                        & (0,0.222)                                    & (0,0.116)                                        \\
Duplicated Lines                         & (0.439,63.284)                                  & (0.145,163.258)                                      & (0.081,124.342)                                  & (0.085, 102.796)                                      \\ \bottomrule
\end{tabular}
\label{tab:metric-parameters2}
\end{table*}

Monotonic metrics, such as 'Code Smells', exhibit an exponential distribution pattern, as represented in Fig. \ref{fig:metric_dist} and Table \ref{tab:metric-parameters2}. This distribution aligns with our understanding that superior code quality is associated with fewer bugs, verifying the effectiveness of our method. Furthermore, the threshold parameter $c$ reflects the tolerance value for full scores. In the probability density function (Eq. \eqref{eq:exp_pdf}) except 'Code Smells', 'Depth Inheritance Tree', and 'Total Violations' where $c$ approximates 1. 

The fitted exponential decay parameter, $\lambda$, reflects the sensitivity of metrics to scores.  Particularly, a $\lambda\lesssim1$ is observed for metrics such as 'File Complexity', 'Depth Inheritance Tree', 'Number of Children', 'Duplicated Blocks', and 'Duplicated Files', which implies a low sensitivity to metric variations of the order of 1. Conversely, the $\lambda$ value for total violation is high, which reflects the high sensitivity of the number of violations.


Non-monotonic metrics, such as the 'Cyclomatic Complexity', follow an asymmetric Gaussian distribution. According to Eq. \eqref{eq:agass_score}, repositories with metric values close to the Gaussian center get higher scores since they fall into the range where high-quality OSS are mostly located. In Table \ref{tab:metric-parameters1}, the Gaussian centers $\mu$ are large ($\gg1$) for the metrics of 'Cyclomatic Complexity', 'Cognitive Complexity', and 'Comment Lines' in most cases except for the 'Comment Lines' of the Javascript and Typescript languages. The latter two distributions are almost monotonic ($\mu=0$), potentially because these two languages are generally easy to understand and do not require additional command lines. 

The fitted widths $\sigma_{1,2}$ are large and have asymmetric sensitivity; i.e. relatively long tails are observed on the right of the asymmetric Gaussian distributions. For "command line" in Python, increasing command lines before the center point has high sensitivity, while it becomes less sensitive after the center point.

After obtaining metric distributions, we score the metrics of each OSS repository based on their respective locations in the distributions.

\subsection{Importance Weights}

Table~\ref{tab:importance_values} shows the feature importance values from the GBC model in Section \ref{sec: application}, which we use as metric score weights in Eq. \eqref{eq: overall_score} within the three dimensions: maintainability, reliability, and functionality. The relative importance values are listed in Table~\ref{tab:importance_values}. We normalized the importance values for each dimension to get relative weights within dimensions. 

\begin{table*}
\small
\centering
\renewcommand{\arraystretch}{1.2}
\caption{Importance Values for Metric Scores}
\begin{tabular}{@{}p{2cm}p{4cm}cccc@{}}
\toprule
\multirow{2}{2cm}{\textbf{Dimension}} & \multicolumn{1}{c}{\multirow{2}{*}{\textbf{Metric}}} & \multicolumn{4}{c}{\textbf{Importance}}                                    \\
  & \multicolumn{1}{c}{}                        & \textbf{Java}           & \textbf{JavaScript}     & \textbf{Python}     & \textbf{TypeScript} \\ \midrule
\multirow{12}{*}{Maintainability}             & Cyclomatic Complexity                       & 0.110 (0.083) & 0.190 (0.082) & 0.250 (0.120) & 0.223 (0.081) \\
  & File Complexity    & 0.220 (0.165) & 0.396 (0.171) & 0.449 (0.215) & 0.402 (0.146) \\
  & Cognitive Complexity                        & 0.086 (0.065) & 0.289 (0.125) & 0.119 (0.057) & 0.215 (0.078) \\
  & Code Smells                                 & 0.066 (0.049) & 0.125 (0.054) & 0.182 (0.087) & 0.160 (0.058) \\
  & Coupling Between Objects                    & 0.096 (0.072) & /              & /              & /              \\
  & Fan-in                                      & 0.108 (0.081) & /              & /              & /              \\
  & Fan-out                                     & 0.057 (0.043) & /              & /              & /              \\
  & Depth Inheritance Tree                      & 0.075 (0.057) & /              & /              & /              \\
  & Number of Children                          & 0.026 (0.020) & /              & /              & /              \\
  & Lack of Cohesion of Methods                 & 0.078 (0.058) & /              & /              & /              \\
  & Tight Class Cohesion                        & 0.010 (0.008) & /              & /              & /              \\
  & Loose Class Cohesion                        & 0.068 (0.051) & /              & /              & /              \\
  & \textbf{Sum}   & 1 (0.752) & 1 (0.432) & 1 (0.479) & 1 (0.363) \\
\midrule
\multirow{3}{*}{Reliability}                  & Total Violations                            & 0.474 (0.056) & 0.288 (0.070) & 0.293 (0.068) & 0.228 (0.065) \\
  & Critical Violations                         & 0.272 (0.032) & 0.420 (0.102) & 0.410 (0.095) & 0.414(0.118) \\
  & Info Violations                             & 0.254 (0.030) & 0.292 (0.071) & 0.297 (0.069) & 0.358 (0.102) \\
  & \textbf{Sum}   & 1 (0.118) & 1 (0.243) & 1 (0.232) & 1 (0.285) \\
\midrule
\multirow{5}{*}{Functionality}                & Line to Cover                               & 0.000 (0.000) & 0.000 (0.000) & 0.000 (0.000) & 0.000 (0.000) \\
  & Comment Lines                               & 0.454 (0.059) & 0.317 (0.103) & 0.370 (0.107) & 0.318 (0.112) \\
  & Duplicated Blocks                           & 0.162 (0.021) & 0.286 (0.093)   & 0.197 (0.057) & 0.148 (0.052) \\
  & Duplicated Files                            & 0.190 (0.025) & 0.120 (0.039) & 0.166 (0.048) & 0.179 (0.063) \\
  & Duplicated Lines                            & 0.194 (0.025) & 0.277 (0.090) & 0.267 (0.077) & 0.355 (0.125) \\ 
  & \textbf{Sum}   & 1 (0.130) & 1 (0.325) & 1 (0.289) & 1 (0.352) \\
  \bottomrule
\end{tabular}
\begin{tablenotes}
    \item  The parenthesis values are original importance values, while the values outside parenthesis are normalized in the dimension level. 
\end{tablenotes}
\label{tab:importance_values}
\end{table*}

In the maintainability dimension, 'File Complexity' has the largest weight across four programming languages, followed by 'Cognitive Complexity' 'Cyclomatic Complexity', and 'Code Smells'. These metrics contribute more to the maintainability scores. For Java repositories, all the coupling and cohesion metrics show similar contributions $\lesssim0.1$, reflecting their weak contribution to OSS adoption.

In the reliability dimension, 'Total Violations' contributes mostly to Java, while 'Critical Violations' contributes mostly to the other three languages, which suggests varying priorities of solving violations for different languages. 

In the functionality dimension, the 'Comment Lines' metric contributes more to Java, potentially because Java is less intuitive to understand, which requires code comments for better understanding. The 'Comment Lines' metric also contributes significantly to the other three scripting languages. We note that zero 'Line to Cover' metric values were obtained in our raw data, either caused by problems in obtaining this metric or because codes in OSS repositories are rarely tested. This gap can be closed when applying our methodology in specific companies where values of 'Line to Cover' are obtained for their close-source repositories.

\subsection{Software Adoption}

\begin{figure*}[t]
  \centering
  {\includegraphics[width=0.4\textwidth]{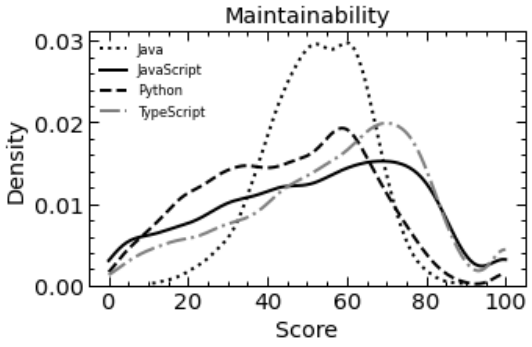}}
  {\includegraphics[width=0.4\textwidth]{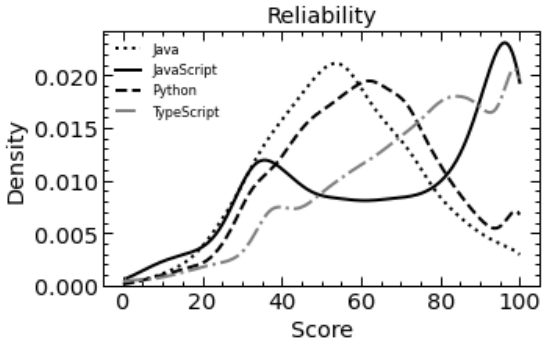}}
  {\includegraphics[width=0.4\textwidth]{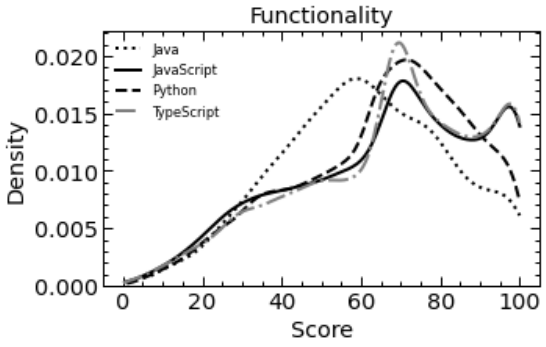}}
  {\includegraphics[width=0.4\textwidth]{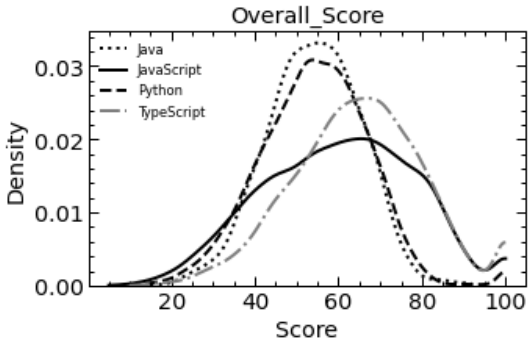}}
  \caption{Overall Scores for Four Languages}
  \label{fig:scores_all}
\end{figure*}

We present the overall scores of included OSS repositories in Fig.~\ref{fig:scores_all} and assess the explanatory power of our metric scores on the OSS stars using Table~\ref{Explainability}. We observe that Java code metric scores show higher explanatory power for the OSS repository's stars compared to the other languages, which suggests that code quality can better determine the success of Java-based OSS repositories in terms of stars received, which may be attributed to the greater availability of metrics for Java or the nature of repositories developed using Java for large-scale platforms and systems.

In contrast, JavaScript, Python, and TypeScript exhibit relatively lower explanatory power of metric scores, indicating their code quality might be less critical in determining their OSS adoption, possibly because of their primary use in data analytics or other domains where their adoption is less influenced by code quality.

\begin{table}
\small
\centering
\caption{Metric Scores Explanatory Power \label{Explainability}}
\begin{tabular}{@{}ccccc@{}}
\toprule
\textbf{Language} & \textbf{Java} & \textbf{JavaScript} & \textbf{Python} & \textbf{TypeScript} \\ \midrule
Accuracy          & 0.947         & 0.826               & 0.808           & 0.817               \\
Precision         & 0.971         & 0.838               & 0.831           & 0.834              \\
Recall            & 0.917         & 0.803               & 0.771           & 0.784               \\
F1                & 0.943         & 0.820               & 0.800           & 0.808               \\
AUC\_ROC          & 0.946         & 0.826               & 0.815           & 0.817               \\
R2                & 0.787    & 0.274 	& 0.186 	& 0.247 \\ 
\bottomrule
\end{tabular}
\end{table}

\section{Conclusion} \label{sec: dis}

Our research focuses on code quality with three dimensions: maintainability, reliability, and functionality. We evaluate metrics based on their distributions. Our study advances the understanding of code quality and contributes to better quality control standards and practices, ultimately supporting the success and sustainability of software. 

Although our study provides valuable implications, it has some limitations that need to be acknowledged. We have not yet systematically validated the effectiveness of the method. Moving forward, it would be beneficial to incorporate validation techniques, such as sensitivity tests, to ensure the accuracy and reliability of the distribution fitting. Additionally, the parameters of the fitted distribution are sensitive to data distribution, making it necessary to incorporate more data for determining them. 

\section*{Acknowledgment}
Y. Xia is partly supported by the "Pioneering Innovator" award from the Guangzhou Tianhe District government. Z. Li is partly supported by the Guangdong Basic and Applied Basic Research Foundation (2021A1515012039). We would like to acknowledge useful discussions and support from Mianmian Zhang and other colleagues at the HSBC Lab.

\singlespacing

\bibliography{main}

\end{document}